\documentclass{ptephy}

\usepackage{amsmath} 
\usepackage{amsthm} 
\usepackage{cite} 
\usepackage{graphicx} 
\usepackage{algorithmic} 
\usepackage{subfig} 
\usepackage{url} 
\usepackage{latexsym}
\usepackage{natbib}
\usepackage{bm}
\bibliography{sample}

\begin{document}
\title{Hubble constants and luminosity distance in the renormalized cosmological models
due to general-relativistic second-order perturbations}
\author{\name{Kenji Tomita}{\ast}}
\address{\affil{}{Yukawa Institute for Theoretical Physics, Kyoto University, Kyoto 606-8502,
 Japan}
\rm{\email{ketomita@yahoo.co.jp}}}

\begin{abstract}
Renormalized cosmological models based on the general-relativistic 
second-order perturbation theory were proposed in the previous papers to
solve a tension on the observed Hubble constants. The cosmological random adiabatic 
fluctuations were found to play an important role as the first-order perturbations. 
The second-order metric
perturbations in a previous paper are revised in the present paper. It is shown
 as a result that two types of Hubble constants (the kinematic constant 
$H_{kin}$ and the dynamic constant $H_{dyn}$) are derived, and their values are
 found to be comparable, and larger than the background value. The optical
quantities such as redshift and luminosity distance are derived using the 
 revised metric perturbations.
\end{abstract}

\maketitle

\section{Introduction}

In order to discuss the cosmological tension on the difference 
between the Hubble constant  derived from the Planck 
measurements\citep{planck1, planck2} and that from
 the direct measurements of the Hubble constant\citep{h1,h11,h2,h3,h4,h5},
 we studied cosmological models\citep{t1,t2,t3}  which were derived using 
 the general-relativistic  second-order perturbation theory (\citep{t01} for
non-zero $\Lambda$ and \citep{t02,russ,mat} for zero $\Lambda$). 
 It was found in these models that the 
cosmological random adiabatic density fluctuations\citep{bbks} play an 
important role as the first-order perturbations for producing the gap of  
Hubble constants due to the non-linear process.

After the publication of our above papers, we found a necessity of the 
revision for the
derivation of averages of second-order metric perturbations in the first 
paper\citep{t1},
 which changed the derived value of the Hubble constant slightly.

In the present paper we first derive our correct averages of metric 
perturbations, and use them to 
derive the Hubble constants and optical quantities such as redshift and
luminosity distance necessary  for the 
observations.
 
In Sect. 2, we show our background model and the outline of our
 perturbation theory. In Sect. 3, we show the revised second-order metric 
 perturbations,
and in Sect. 4, derive two kinds of the Hubble parameter, which are found to 
be comparable, and larger than the background value.
In Sect. 5, we derive the optical quantities, and the observational relation
using the revised metric perturbations. In Sect. 6, concluding remarks are 
given. In Appendix A, the basic formulation is compactly reviewed. In 
Appendix B,  the deceleration parameter $q$ is derived, and in Appendix C, the 
luminosity distance $d_L$ is derived, based on the revised metric 
perturbations. In Appendix D, dependence of renormalized model parameters on 
background model parameters is shown.
 
\section{Background and the perturbation theory}

The space-time of our spatially flat background universe is expressed by the 
line element
\begin{equation}
  \label{eq:a1}
ds^2 = g_{\mu \nu} dx^\mu dy^\nu = a^2 (\eta) [-d\eta^2 + \delta_{ij} 
dx^i dx^j ],
\end{equation}
where the Greek and Roman letters denote $0, 1, 2, 3$ and $1, 2, 3$,
 respectively. The conformal time $\eta (= x^0)$ is related to the cosmic
 time $t$ by $dt = a(\eta) d\eta$. The background Hubble parameter $H \
 (\equiv a'/a^2 = \dot{a}/a)$ satisfies
\begin{equation}
  \label{eq:a2a}
H \ = \  [(\rho + \Lambda)/3]^{1/2} \ = \ H_0 \ (\Omega_M a^{-3} +
 \Omega_\Lambda)^{1/2}, 
\end{equation}
where  a prime and a dot denote $d/d \eta$ and $d/d t$, respectively. 
We use a background model with  model parameters given by
\begin{equation}
  \label{eq:a2b}
H_0 = 67.3 \ {\rm km \ s^{-1} Mpc^{-1}} \quad  {\rm{and}} \quad
(\Omega_M,  \Omega_{\Lambda} )  = (0.22, 0.78),
\end{equation}
where
\begin{equation}
  \label{eq:a3}
\Omega_M  = \frac{8\pi G \rho_0}{3H_0^2} = \frac{1}{3} \frac{\rho_0}{H_0^2}
  \quad {\rm{and}} \quad \Omega_{\Lambda} = \frac{\Lambda c^2}{3H_0^2} = 
  \frac{1}{3}   \frac{\Lambda}{H_0^2},
\end{equation}
$H_0$ and $\rho_0$ are the present Hubble constant and  matter density,
 and the units $8\pi G = c = 1$ are used. In Appendix D,  cases of 
$ (\Omega_M,  \Omega_{\Lambda} )  = (0.24, 0.76)$ and $(0.28, 0.72)$ are 
treated for comparison. 

For perturbations on large scales with $x \equiv k/k_{eq} \leq x_{max}$,
the perturbed metric, velocity and density perturbations are expressed as
\begin{equation}
  \label{eq:a4}
   \begin{split}
\delta g_{\mu\nu}  \ &= \ h_{\mu\nu} \ + \ \ell_{\mu\nu}, \\
\delta u^\mu \ &= \ \mathop{\delta}_1 u^\mu \ + \ \mathop{\delta}_2 u^\mu, \\ 
\delta \rho/\rho \ &= \ \mathop{\delta}_1 \rho/\rho \ + \ \mathop{\delta}_2 
\rho/\rho,
 \end{split}
\end{equation}
where the definition of $k_{eq}$ and $x_{max}$ are shown in Appendix A.
Here we assume the synchronous and comoving coordinates, that is
\begin{equation}
  \label{eq:a5}
h_{00} = 0,  \ h_{0i} = 0 \ \ {\rm and} \ \ \delta_1u^0 = 0, \ \delta_1 u^i = 0, 
\end{equation}
\begin{equation}
  \label{eq:a6}
\ell_{00} = 0, \ \ell_{0i} = 0 \ \ {\rm and} \ \ \delta_2 u^0 = 0, \ \delta_2 u^i = 0
\end{equation}
in  the same way as the previous paper\citep{t1}, cited as [I].  
In the previous paper\citep{t01}, the expressions of metric in the Poisson 
coordinates also were shown, but here our treatments are confined only to the 
synchronous and comoving coordinates. 
  
The first-order perturbations in the growing mode are expressed in Eq.(14)
 of [I]  by use of an arbitrary potential function $F({\bf x})$, where 
 ${\bf x}$ is the spatial coordinates. The amplitude of $F({\bf x})$ is related 
 to the cosmological adiabatic density fluctuations.\citep{bbks} 
 The  second-order perturbations
corresponding to the first-order perturbations are expressed  in Eqs. (20)
- (23) of [I].  
 
The average values of second-order density perturbations  are shown in  
Appendix A with some small corrections. For those of second-order metric
 perturbations, the revised version is shown in the next section.
 
The scale with $x > x_{max}$ represents the scale which is always sub-horizon 
at the matter-dominant stage after the epoch such as $1 + z = 1500$ \ (cf [10]).
Perturbations on small scales with $x \geq x_{max}$ were separately treated 
in the Newtonian approximation and their effect to the large-scale quantities 
was found to be negligible. So the following analyses are confined to the 
above perturbations on large scales with $x \leq x_{max}$. 
  
\section{Revised second-order metric perturbations}

The average of second-order perturbations of the scale-factor ($\delta_2 a$) is
 expressed using  the second-order metric perturbations $l_{ij}$ as
\begin{equation}
  \label{eq:b1}
\delta_2 (a^2) / a^2 \quad = \quad \frac{1}{3} \langle l^m_m  \rangle,
 \end{equation} 
where the average process is shown in Appendix A and  $\langle l^m_m  \rangle
 = \langle l^1_1 + l^2_2 + l^3_3  \rangle $. We have this relation 
(\ref{eq:b1}), because  $a^2 l^i_j$ represents the perturbations corresponding to the
 background space-time Eq.(1). So, Eq.(40) of [I] is wrong with respect to 
the factor $a^2$, 
 and the following Eqs. (42) - (51) of [I] should be replaced by the 
 correct ones, which are shown in the following:
\begin{equation}
  \label{eq:b2}
\langle l_{ij}  \rangle \ = \ P(\eta) \langle L_{ij}  \rangle + P^2 (\eta) \langle M_{ij} 
 \rangle
+ Q(\eta) \langle N_{,ij}  \rangle + \langle C_{ij}  \rangle.
 \end{equation} 
Here $L_{ij}, M_{ij}$ and $N_{,ij}$ are metric components being functions of spatial variable 
${\bf x}$, and $C_{ij}$ represents the components of gravitational waves, which were
used in Eqs. ((20), (21), and (A1) - (A4)) of [I].
Since $L^i_j = L_{ij}, \ M^i_j = M_{ij}$ and 
\begin{equation}
  \label{eq:b3}
    \begin{split}
L^i_i &= - \frac{1}{2} \ [2 F\Delta F + \frac{3}{2} F_{,l} F_{,l}], \\
M^i_i &= \frac{1}{28} \ [10 F_{,jl} F_{,jl} - 3 (\Delta F)^2 ],  \\
\Delta N &=  \frac{1}{28} \ [ (\Delta F)^2 -  F_{,kl} F_{,kl}], \\
\Box C^i_i &= 0,
\end{split}
 \end{equation} 
we obtain
\begin{equation}
  \label{eq:b4}
    \begin{split}
\langle L^i_i  \rangle &= - \frac{1}{4} \ \langle F \Delta F  \rangle, \\
\langle M^i_i  \rangle &=  \frac{1}{4} \ \langle (\Delta F)^2  \rangle, \\
\langle \Delta N  \rangle &= \langle C^i_i  \rangle  = 0.
\end{split}
 \end{equation} 
Then we get using Eqs.(\ref{eq:e5}) and (\ref{eq:e7})
\begin{equation}
  \label{eq:b5a}
\langle l_{ii}  \rangle =  \langle l^i_i  \rangle  = \frac{1}{4} (2\pi)^{-2} P(\eta) 
\Big[ \int d{\bf k} k^2 
 {\cal P}_F ({\bf k}) + P(\eta)  \int  d {\bf k} k^4 {\cal P}_F ({\bf k})\Big].
 \end{equation} 
Here ${\cal P}_F$ is replaced by ${\cal P}_R$ with $ {\cal P}_{R0}$ in Eq.(\ref{eq:e10}), 
and we obtain
\begin{equation}
  \label{eq:b5}
\langle l_{ii}  \rangle  =  2\pi \ 32.4^4 \ {\cal P}_{R0} 
 \  Z(a) [32.4^{-2} A + Z(a) B], 
 \end{equation} 
where \ $P(\eta)$ is expressed using $Z(a)$ (defined in Eqs. (\ref{eq:e14}) 
and (\ref{eq:e15})), and  the constants $A$ and $B$ \ (defined  by  
Eq. (\ref{eq:e13})) reflect the amplitude of the BBKS adiabatic fluctuations.\citet{bbks}. 

The average metric perturbations $\langle l_{ii}  \rangle$ are spatially constant and 
isotropic, and so we can consider  the renormalized scale-factor $a_{rem}$ defined by
\begin{equation}
  \label{eq:b6}
a_{rem} = a \ \Bigl(1 + \frac{1}{3} \langle l_{ii} \rangle \Bigr)^{1/2} = a \ \Bigl(1 + 
\frac{1}{6} \langle l_{ii} \rangle \Bigr),
 \end{equation} 
where we neglect the terms of higher-orders than  second-order. The renormalized
 Hubble parameter ($H_{kin}$) is defined as 
\begin{equation}
  \label{eq:b7}
H_{kin} \equiv \frac{\dot{a}_{rem}}{a_{rem}} = \frac{\dot{a}}{a} + \frac{1}{6} 
{\langle l_{ii} \rangle}^.
 \end{equation} 
or 
\begin{equation}
  \label{eq:b8}
H_{kin} /H = 1 + \frac{1}{6}  {\langle l_{ii} \rangle}^. /H,
 \end{equation} 
where $H\ (\equiv \dot{a}/a)$ is the background Hubble parameter.
Here $H_{kin}$ denotes the kinematic definition of the Hubble parameter.
Differentiating Eq.(\ref{eq:b5}), we obtain
\begin{equation}
  \label{eq:b9}
{\langle l_{ii}  \rangle}^.   =  2\pi \ 32.4^4 \ {\cal P}_{R0} 
 \ [32.4^{-2} A + 2 Z(a) B] \ \frac{d Z(a)}{da} \dot{a}, 
 \end{equation} 
where 
\begin{equation}
  \label{eq:b10}
\frac{d \ Z(a)}{da} = (H_0)^2 \frac{P'}{a'} = (H_0/H)^2 \ Y(a)/a^3 = 
\frac{Y(a)}{\Omega_M + \Omega_\Lambda a^3}
 \end{equation} 
using Eqs. (\ref{eq:a2a}) and (\ref{eq:e15}). Therefore, we get using Eq. (\ref{eq:b10})  
\begin{equation}
  \label{eq:b11}
{\langle l_{ii}  \rangle}^. /H   =  2\pi \ 32.4^4 \ {\cal P}_{R0} 
 \frac{ [32.4^{-2} A + 2 Z(a) B] \ Y(a) a }{\Omega_M + \Omega_\Lambda a^3},
 \end{equation} 
and then the kinematic second-order Hubble parameter is given by
Eq.(\ref{eq:b7}).   At present epoch, it is expressed  as
\begin{equation}
  \label{eq:b12}
(H_{kin})_0 /H_0 = 1 + \frac{1}{6}  [{\langle l_{ii} \rangle}^. /H]_0
=  1 + \frac{2\pi}{3} \ 32.4^4 \ {\cal P}_{R0} 
 \ \Bigl[\frac{1}{2} \times 32.4^{-2} A + Z(1) B \Bigr] \ Y(1) .
 \end{equation} 

Here, \ $Y(1)$ and $Z(1)$ are expressed as
\begin{equation}
  \label{eq:b13}
Y(1) = I(1),  \quad Z(1) = \frac{2}{3\Omega_M} [1 -  I(1)], \quad
{\rm and} \quad I(1) = \int^1_0 db [b^3/(\Omega_M + \Omega_\Lambda b^3)]^{1/2}.
\end{equation}
For the background model parameters (\ref{eq:a2b}), we get
\begin{equation}
  \label{eq:b14}
I(1) = Y(1) = 0.566,  \quad Z(1) = 1.316.
\end{equation}
So, we obtain using $A$ and $B$ in Eq.(\ref{eq:e18})
\begin{equation}
  \label{eq:15}
(H_{kin})_0 = 72.6 \ {\rm km \ s^{-1} Mpc^{-1}} .
 \end{equation} 

In order to explain the histories of $H_{kin}$ and $a_{rem}$, the behaviors 
of $H_{kin} \ (H_0/H)$ and $\xi \ (\equiv a_{rem} (t)/a(t) - 1)$  are 
shown as functions of $a$ in Figs. 1 and 2, respectively. Corresponding 
previous figures (Figs. 2 and 4 in [I]) must be replaced by these figures.  

\begin{figure}[t]
\caption{\label{fig:1} History of $H_{kin}$. The ordinate denotes $H_{kin}
(H_0/H)$ is expressed as a function of $a$. The scale factor $a$ has $1$ 
at the present epoch.} 
\centerline{\includegraphics[width=10cm]{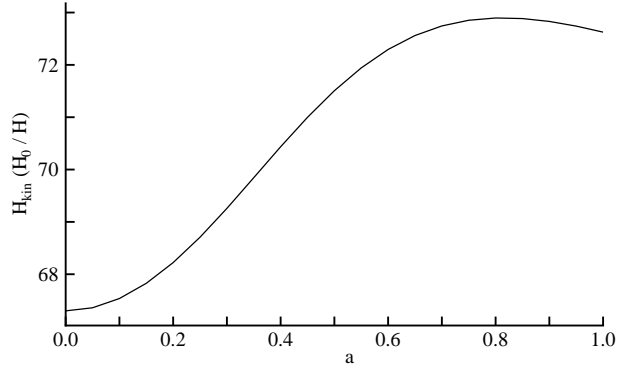}}
\end{figure}
\begin{figure}[t]
\caption{\label{fig:2} The relative scale-factors $\xi \ (\equiv a_{rem}/a - 1)$
is expressed as a function of $a$. The scale factor $a$ has $1$ at the present
 epoch.} 
\centerline{\includegraphics[width=10cm]{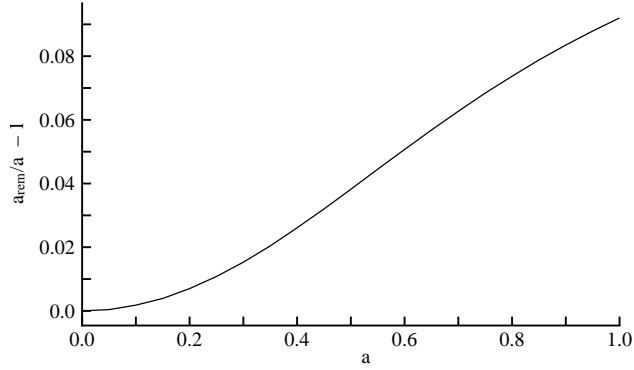}}
\end{figure}
 
\section{Renormalization of model parameters}
 
For the second-order density perturbations, we have no revision, and so  
using Eqs.(\ref{eq:e12}) and (\ref{eq:e19}) for the density perturbations, 
we can consider the renormalization of model parameters, similarly to that in [I].  
Since $\langle \delta_2 \rho \rangle$ is spatially constant and isotropic, 
we assume that it is a part of the renormalized matter density $\rho_{rem}$.
So we have
\begin{equation}
  \label{eq:c1}
\rho_{rem} = \rho \ + \ \langle \delta_2 \rho  \rangle.
 \end{equation} 
Then the renormalized ones corresponding  to  $\Omega_M $ and
 $\Omega_\Lambda$ are given by
\begin{equation}
  \label{eq:c2}
[(\Omega_M)_{rem}, (\Omega_\Lambda)_{rem}] = [\Omega_M 
(1 + \langle \delta_2 \rho/\rho \rangle), \quad \Omega_\Lambda]/(1 + \langle
 \delta_2 \rho/\tilde{\rho} \rangle),
 \end{equation} 
where $\tilde{\rho} \equiv \rho + \Lambda$.

Next, we define a renormalized Hubble parameter $H_{dyn}$ \
 corresponding to Eq.(\ref{eq:c1}) as
\begin{equation}
  \label{eq:c3}
H_{dyn} = \Bigl[\frac{1}{3} (\rho_{rem} + \Lambda) \Bigr]^{1/2} = 
\Bigl[\frac{1}{3} (\tilde{\rho} + \langle \delta_2 \rho \rangle) \Bigr]^{1/2}.
 \end{equation} 
This Hubble parameter appears when we describe the dynamical evolution
of the perturbed model, as used in the previous paper\citep{t2}, and so 
 we call it the dynamical Hubble parameter. Its present value is expressed as
\begin{equation}
  \label{eq:c4}
(H_{dyn})_0 /H_0 =  1 + \frac{2\pi}{3} \ 32.4^4 \ {\cal P}_{R0} 
 \ \Bigl[-\frac{5}{2} \times 32.4^{-2} A + Z(1) B \Bigr]  [1 - Y(1)] ,
 \end{equation} 
 where $A$ and $B$ are given in Eqs.(\ref{eq:e13}) and (\ref{eq:e18}).  
 
 Corresponding to the background model parameter (\ref{eq:a2b}), the 
 present value of renormalized model parameters are found to be 
\begin{equation}
  \label{eq:c5}
(\Omega_M)_{rem} = 0.305, \quad  (\Omega_\Lambda)_{rem} = 0.695,
 \end{equation} 
and 
\begin{equation}
  \label{eq:c6}
(H_{dyn})_0 = 71.4 \ {\rm km \ s^{-1} Mpc^{-1}} .
 \end{equation} 
As for the histories of $\langle \delta_2 \rho \rangle$ and $(\Omega_M)_{rem}$,
Fig. 1 and Fig. 3 in [I] are useful also in the present paper.

On the other hand,  we have the kinematic Hubble parameter $H_{kin}$,
 which was derived in the previous section. 
$H_{kin}$ and $H_{dyn}$ are different with respect to the factor $Y(a)$, and 
$H_{kin}$ is a little larger than $H_{dyn}$.  So we discriminate these two Hubble 
parameters.  On the other hand, both Hubble parameters are found to be 
larger than the background Hubble parameter $H$.

These Hubble parameters depend on the value of $B$
which is sensitively related to the upper limit $x_{max}$ in Eq.(\ref{eq:e13}). 
The terms with $A$ are negligibly small.
We have used $x_{max} = 5.7$ here and in previous papers.
In Table \ref{table 1}, we show the dependence of $(H_{kin})_0$ and
$(H_{dyn})_0$ on the value of 
$x_{max}$ and $L_{max} \ (= 2\pi/k_{max})$ given in Eq.(53) of [I],
which may represent the boundary for whether the general-relativistic
 non-linearity is effective  for the evolution of perturbations, as was discussed
  in a previous paper\citep{t3}.  For larger $x_{max}$ (or smaller $L_{max}$), we have
  larger Hubble constant.    
 

 \begin{table}[!h]
\caption{Dependence of $(H_{kin})_0$ and  $(H_{dyn})_0$ on $x_{max}$ and $L_{max}$ \
in the case of  background model parameters (\ref{eq:a2b}).}
\label{table 1}
\centering
\begin{tabular}{cccc}
\hline
$x_{max}$ & $L_{max}$ & $(H_{kin})_0$ & $(H_{dyn})_0$ \\ 
\hline
$5.7$ & $102/h$ & $72.6$ & $71.4$ \\ 
$6.0$ & $97/h$ & $73.3$ & $71.7$ \\
$6.3$ & $92/h$ & $74.0$ & $72.4$ \\
\hline
\end{tabular}
\end{table}
 
 \medskip 

\section{Optics and observations}

The renormalized line-element can be expressed as
\begin{equation}
  \label{eq:d1}
ds^2 = - dt^2 \ + \  {a_{rem} (t)}^2 \ [(dx^1)^2 + (dx^2)^2 + (dx^3)^2],
 \end{equation} 
 where the renormalized scale factor $a_{rem} (t)$ is given by 
 Eq. (\ref{eq:b6}), and $x^1, x^2$ and $x^3$ are comoving coordinates.
 
\subsection{Redshift}
 The light path is given by the null condition
\begin{equation}
  \label{eq:d2}
dt = \pm  {a_{rem} (t)} dr, 
 \end{equation} 
where $ r \equiv [(dx^1)^2 + (dx^2)^2 + (dx^3)^2]^{1/2}$.

  If a light ray starts from a distant source with $r$ at epoch $t_1$
and reaches an observer at epoch $t_0$, we have
\begin{equation}
  \label{eq:d3}
\int^{t_0}_{t_1} dt/a_{rem} = r. 
 \end{equation} 
If we receive two subsequent signals with intervals $\delta t_0$ and 
$\delta t_1$ from a comoving source, we obtain
\begin{equation}
  \label{eq:d4}
\frac{\delta t_1}{a_{rem} (t_1)} \ = \ \frac{\delta t_0}{a_{rem} (t_0)} .
 \end{equation} 
For the frequencies $\nu_0$ and $\nu_1$ \ (given by $\nu_1/\nu_0 =
\delta t_0/\delta t_1$),
we have 
\begin{equation}
  \label{eq:d5}
1 + z_{rem} \ = \ \nu_1 /\nu_0 \ = \ a_{rem} (t_0)/ a_{rem} (t_1),
 \end{equation} 
where $z_{rem}$ is the redshift.

 For neaby sources, we can expand $a_{rem} (t)$ as
\begin{equation}
  \label{eq:d6}
a_{rem} (t) =  a_{rem} (t_0) [1 + (t - t_0) (H_{kin})_0 + \cdot \cdot \cdot],
 \end{equation} 
where $(H_{kin})_0$ gives the relation
\begin{equation}
  \label{eq:d7}
(H_{kin})_0 = \dot{a}_{rem}(t_0) / a_{rem} (t_0).
 \end{equation} 

Moreover, we have
\begin{equation}
  \label{eq:d8}
z_{rem} = (H_{kin})_0 \Delta t, .
 \end{equation} 
where $\Delta t = t_0 - t_1$.

\subsection{Luminosity distance}

The relation between the apparent luminosity $l$ and the absolute 
luminosity $L$ is expressed by
\begin{equation}
  \label{eq:d9}
l = L/(4 \pi {d_L}^2), 
 \end{equation} 
where $d_L$ is the luminosity distance between a source and an observer.
In the expanding universe with the metric (\ref{eq:d1}), the time intervals
$\delta t_1$ and $\delta t_0$ in the source and the observer are not equal and
given by Eq.(\ref{eq:d4}).  Moreover, the received and emitted energies of a
 photon are different and given by the redshift factor. As a result, the above
  relation is expressed as
\begin{equation}
  \label{eq:d10}
l = L/[4 \pi (d_L)^2_{rem} ],
 \end{equation} 
using the renormalized luminosity distance $(d_L)_{rem}$, which is
given by
\begin{equation}
  \label{eq:d11}
(d_L)_{rem} = a_{rem} (t_0) \ r(z_{rem}) \ (1 + z_{rem}).
 \end{equation} 
Here $r(z_{rem})$ is the coordinate distance between the observer and 
the source with $z_{rem}$, which is derived eliminating $t_1$ from Eqs.
(\ref{eq:d3}) and (\ref{eq:d5}).

For $z_{rem}  \ll 1$, we have
\begin{equation}
  \label{eq:d13}
z_{rem} = (H_{kin})_0 (t_0 - t_1)  + \frac{1}{2} [(q_{kin})_0 +2] \ (H_{kin})^2_0 \
 (t_0 - t_1)^2 + \cdot \cdot \cdot,
 \end{equation} 
where the kinematic deceleration parameter $(q_{kin})$ is defined as
\begin{equation}
  \label{eq:d14}
q_{kin} \ \equiv \ - \Bigl(\frac{d^2 a_{rem} (t)}{dt^2} \Bigr)/[(H_{kin})^2 \ a_{rem}].
 \end{equation} 
From Eq.(\ref{eq:d13}), we obtain inversely
\begin{equation}
  \label{eq:d15}
(H_{kin})_0 (t_0 - t_1) = z_{rem}  - \frac{1}{2} [ (q_{kin})_0 +2] \ z_{rem}^2 +
 \cdot \cdot \cdot 
 \end{equation} 
Therefore, we obtain from Eq.(\ref{eq:d11})
\begin{equation}
  \label{eq:d16}
(d_L)_{rem} = (H_{kin})_0^{-1} \Bigl\{z_{rem} + \frac{1}{2} [1 - (q_{kin})_0] \ z_{rem}^2 
\ + \  \cdot \cdot \cdot \ \Bigr\}, 
 \end{equation} 
where the value of $q_{kin}$ is derived  in Appendix B.  Its present value
$(q_{kin})_0$ is a little larger than $q_0$ in the background :
\begin{equation}
  \label{eq:d17}
q_0 = -0.670 \quad {\rm and} \quad (q_{kin})_0 = - 0.659
 \end{equation} 
for the background model parameter (\ref{eq:a2b}). 

The background equation corresponding to Eq.(\ref{eq:d16}) is\citep{wein}
\begin{equation}
  \label{eq:d17a}
d_L = (H_0)^{-1} \Bigl\{z + \frac{1}{2} [1 - q_0] \ z^2 
\ + \  \cdot \cdot \cdot \Bigl\}.
 \end{equation} 
Here the ratio $[(q_{kin})_0/ q_0]^{-1} \ (=
1.016)$ is smaller than the ratio $(H_{kin})_0/H_0 \ (=1.079)$.
As for coefficients of the second terms ($\frac{1}{2} [1 - (q_{kin})_0]$
and  $\frac{1}{2} [1 - q_0]$ for $d_L$) also, the ratio is
$(0.993)^{-1} \ (= 1.007)$, and so smaller than $(H_{kin})_0/H_0$.
 
Now let us show the $z_{rem}$-dependence of $(d_L)_{rem}$ from the definition 
(Eq.(\ref{eq:d11})) for arbitrary $z_{rem}$:
\begin{equation}
  \label{eq:d18}
 \begin{split}  
(H_{kin})_0 \ & (d_L)_{rem} = z_{rem} (1 + z_{rem}) \  \Phi (z_{rem}) \\
&\times \frac{1}{z} \int^1_{1/(1+z)}
 \frac{da}{a^2} (\Omega_M a^{-3} + \Omega_\Lambda)^{-1/2} 
 \ \Bigl\{1 + \frac{1}{6} \zeta [(Z(1))^2 - (Z(a))^2] \Bigr\},
 \end{split}
 \end{equation} 
where $\zeta \equiv 2 \pi \times 32.4^4 \ {\cal P}_{R0} B \ (= 0.319)$, \ 
$Z(a)$ is given by Eq.(\ref{eq:e15}), and
\begin{equation}
  \label{eq:d18a}
\Phi (z_{rem}) \equiv \frac{(H_{kin})_0}{H_0} \frac{z}{z_{rem}}.  
 \end{equation} 

The derivation of Eq.(\ref{eq:d18}) is shown in Appendix C.
On the other hand, we have the relation between $z_{rem}$ and $z \ (=
1/a \ - 1)$:
\begin{equation}
  \label{eq:d19}
z_{rem}  = z \ + \ \frac{1}{6} \zeta \{[Z(1)]^2 - [Z(a)]^2 \},
 \end{equation} 
which is derived from Eq.(\ref{eq:g6}). 
So we can get $(H_{kin})_0 (d_L)_{rem}$ as a function of given $z_{rem}$.
The expanded form of this $(d_L)_{rem}$ for small $z_{rem}$ is found to be
 consistent with Eq.(\ref{eq:d16}),  in which the approximate forms of $W$ and
 $\Phi$ in Eqs.(\ref{eq:g11}) and  (\ref{eq:g12}) are used.

On the other hand, the background equation corresponding to 
Eq.(\ref{eq:d18}) is\citep{wein} 
\begin{equation}
  \label{eq:d20}  
H_0 \ d_L =  (1 + z) \times  \int^1_{1/(1+z)}
 \frac{da}{a^2} (\Omega_M a^{-3} + \Omega_\Lambda)^{-1/2} .
 \end{equation} 

The difference between $(H_{kin})_0 (d_L)_{rem}$ and $H_0 d_L$ for equal
$z_{obs}$ is found to be $\sim 0.2 \%$, for $z_{obs}  = 0.5$. 
But,  $(d_L)_{rem}$ and $d_L$ have a larger difference  ( $\sim 8 \%$) due to the ratio 
$(H_{kin})_0 / H_0$.
Here $z_{obs}$ is defined to be equal to $z_{rem}$ and $z$ for the renormalized case
 and the background case, respectively.

The relation between $\log_{10} \ (d_L)_{rem}$ and the observed redshift 
$z_{obs}$ is shown in Fig. 3 for the cases of $z_{obs}  \ <  \ 1$,  in a comparison  
with the background counterpart ($\log_{10} \ d_L$). 
 The absolute magnitude $M$ of an object is defined in terms of an apparent
  magnitude $m$ and the luminosity distance ($(d_L)_{rem}$) as
\begin{equation}
  \label{eq:d21}  
m - M = 5 \ \log_{10} \ [(d_L)_{rem}/ 10 {\rm pc} ],
 \end{equation} 
which is applicable to the observational redshft-magnitude relation for 
SNIa\citep{h1}.  
 
\begin{figure}[t]
\caption{\label{fig:3} Luminosity distances. Solid and dotted curves show
 $ (d_L)_{rem}$ and \ $ d_L$, respectively.  relative to
the observed redshift $z_{obs}$, where $z_{obs}$ is equal to $z_{rem}$ and $z$,
 respectively.} 
\centerline{\includegraphics[width=10cm]{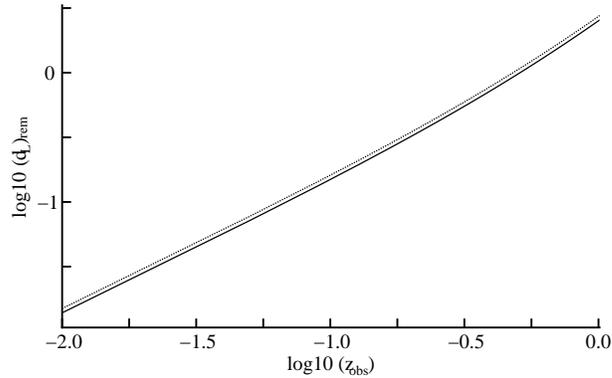}}
\end{figure}

\subsection{Angular diameter distance}

The angular diameter distance $d_A$ is related to $d_L$ as\citep{wein,tsuji}
\begin{equation}
  \label{eq:d12a}
d_A = (1 + z)^{-2} \ d_L .
 \end{equation} 
On the other hand, the angular diameter distance $(d_A)_{rem}$ is related 
to $(d_L)_{rem}$ in the line-element (\ref{eq:d1}) as
\begin{equation}
  \label{eq:d12b}
(d_A)_{rem} = (1 + z_{rem})^{-2} \ (d_L)_{rem} ,
 \end{equation} 
where $z_{rem}$ is related to $z$ by Eq.(\ref{eq:d19}). So, the difference of 
$(H_{kin})_0 (d_A)_{rem}$ and $H_0 d_A$ is very small, for equal $z_{obs}$,
similarly to that of the luminosity distance. 

\section{Concluding remarks}

It was shown that there are two kinds of renormalized Hubble parameters :
the dynamical parameter ($H_{dyn}$) and the kinematic parameter ($H_{kin}$).
As for their present values, the ratio is
\begin{equation}
  \label{eq:d22}
H_0  \ : \ (H_{dyn})_0 \ : \ (H_{kin})_0 \ = \ 1 \ : \ 1.061\ : \ 1.079, 
\end{equation}
so that $(H_{dyn})_0$ and $(H_{kin})_0$ are larger than the background 
constant $H_0$ by the factors $6 \sim 8 \%$, respectively.

The roles of $H_{dyn}$ and $H_{kin}$ are for dynamical motions (including 
phenomena treated in the second paper\citep{t2}) and the optical phenomena
(which were treated in the present paper), respectively. In the latter, we 
 found that $H_0 d_L (z_{obs})$ and $(H_{kin})_0 (d_L)_{rem} (z_{obs})$ are almost
 equal, so that $d_L (z_{obs})$ and $(d_L)_{rem} (z_{obs})$ are different by
 the factor $(H_{kin}/ H_0)$. This situation is quite same also in that of the
 angular diameter distances $d_A$ and $(d_A)_{rem}$ in Eqs.(\ref{eq:d12a}) and 
 (\ref{eq:d12b}).
 
 In the determination of the Hubble constant due to the gravitational-wave
 measurements\citep{abb} also, we have the kinematic constant $(H_{kin})_0$
 in the same way as the optical observation.

\bigskip

\appendix
\section{Average second-order perturbations}

The arbitrary potential function is given the following expression 
\begin{equation}
  \label{eq:e1}
F({\bf x})  = \int d{\bf k} \ \alpha ({\bf k}) \ e^{i{\bf kx}},
\end{equation}
where $\alpha ({\bf k}) $ is a random variable and the average of $F$
 expressed as $\langle F \rangle $ vanishes, and the average of their
 products is given by 
\begin{equation}
  \label{eq:e2}
\langle \alpha ({\bf k}) \alpha ({\bf k'}) \rangle = (2\pi)^{-2}
{\cal P}_F ({\bf k}) \delta ({\bf k} + {\bf k'}).
\end{equation}
Here we have 

\begin{equation}
  \label{eq:e3}
\langle \delta_1 \rho/ \rho \rangle = 0
\end{equation}
for the first-order density perturbation. For the second-order  perturbations,
we have
\begin{equation}
  \label{eq:e4}
    \begin{split}
\langle F_{,i}  F_{,i} \rangle &= - \int \int d{\bf k}  d{\bf k'} 
\langle \alpha ({\bf k}) \alpha ({\bf k'}) \rangle {\bf k} {\bf k'} e^{i({\bf k}+
{\bf k'}) {\bf x}}, \\
\langle F \Delta F \rangle &= - \int \int d{\bf k}  d{\bf k'} 
\langle \alpha ({\bf k}) \alpha ({\bf k'}) \rangle k^2 e^{i({\bf k}+
{\bf k'}) {\bf x}},
  \end{split}
\end{equation}
so that we obtain
\begin{equation}
  \label{eq:e5}
\langle F_{,i}  F_{,i} \rangle  =  - \langle F \Delta F \rangle = (2\pi)^{-2}  
\int d{\bf k} \ k^2 {\cal P}_F ({\bf k}),
\end{equation}
where we corrected some careless misprints in [I]. 

Similarly, we have
\begin{equation}
  \label{eq:e6}
    \begin{split}
\langle F_{,ij}  F_{,ij} \rangle &=  \int \int d{\bf k}  d{\bf k'} 
\langle \alpha ({\bf k}) \alpha ({\bf k'}) \rangle ({\bf k} {\bf k'})^2 e^{i({\bf k}+
{\bf k'}) {\bf x}}, \\
\langle (\Delta F)^2 \rangle &=  \int \int d{\bf k}  d{\bf k'} 
\langle \alpha ({\bf k}) \alpha ({\bf k'}) \rangle k^2 (k')^2 e^{i({\bf k}+
{\bf k'}) {\bf x}},
  \end{split}
\end{equation}
so that we obtain
\begin{equation}
  \label{eq:e7}
\langle F_{,ij}  F_{,ij} \rangle  =  \langle (\Delta F)^2 \rangle = (2\pi)^{-2} 
 \int d{\bf k} \ k^4 {\cal P}_F ({\bf k}).
\end{equation}
The second-order density perturbations are expressed by Eq.(23) of [I] using 
$F ({\bf k})$, and so average  second-order density perturbations are shown 
as follows:
\begin{equation}
  \label{eq:e8}
\langle  \mathop{\delta}_2 \rho/\rho \rangle = 
\frac{1 - \frac{a'} {a}P'}{2\rho a^2} (2\pi)^{-2} \Big[-\frac{5}{2} \int d{\bf k} k^2 
 {\cal P}_F ({\bf k}) + P \int  d {\bf k} k^4 {\cal P}_F ({\bf k})\Big].
\end{equation}
Here $F$ is related to the curvature fluctuation ${\cal R}$ by $F = 2\ {\cal R}$,
and so we have the relation 
\begin{equation}
  \label{eq:e9}
 {\cal P}_F  ({\bf k}) = 4\  {\cal P}_{\cal R}  ({\bf k}),
\end{equation}
where ${\cal P}_{\cal R}$ is  expressed using the power spectrum\citep{wein,
 tsuji}  as
\begin{equation}
  \label{eq:e10}
 {\cal P}_{\cal R} = 2 \pi^2 \  {\cal P}_{{\cal R}0} \ k^{-3} (k/k_{eq})^{n-1} \
  T_s^2  (k/k_{eq})
\end{equation}
and $ {\cal P}_{{\cal R}0} = 2.2\times 10^{-9}$ according to the result of Planck
measurements.\citep{planck1,planck2}
The transfer function $T_s (x)$ is expressed as a function of $x = k/k_{eq}$, 
where 
\begin{equation}
  \label{eq:e11}
k_{eq} \ (\equiv a_{eq} H_{eq}) \ = \ 219 \ (\Omega_M h) \ H_0 \ = \ 32.4 \ H_0. 
\end{equation}
Here $H_0$ \ ($\equiv 100h)$ \ is the present
 background  Hubble constant, \ $(a_{eq}, H_{eq})$ is $(a, H)$ at the epoch of
  equal energy density, and $(\Omega_M, h) = (0.22, 0.673)$ \ (given in
  Eq. (\ref{eq:a2b})).

Moreover, we assume $n = 1$ here and in the following. 
Then we obtain for arbitrary $a$ 
\begin{equation}
  \label{eq:e12}
\langle  \mathop{\delta}_2 \rho/\tilde{\rho} \rangle = \frac{4\pi}{3} \ 32.4^4 \
 {\cal P}_{{\cal R}0} \ \frac{[1 - Y(a)]}{(\Omega_M/a + 
 \Omega_\Lambda a^2)} \Big[- \frac{5}{2} \ 32.4^{-2} A + Z(a) B\Big], 
\end{equation}
where $\tilde{\rho} \equiv \rho + \Lambda$,  and $A$ and $B$ are expressed
 as 
\begin{equation}
  \label{eq:e13}
 A \equiv \int^{x_{max}}_{x_{min}} dx \ x \ T_s^2 (x),  \quad   B \equiv
  \int^{x_{max}}_{x_{min}}
 dx \ x^3 \ T_s^2 (x)
\end{equation}
using the transfer function $T_s (x)$ for the interval ($x_{max}, x_{min}$).  
Here we have
\begin{equation}
  \label{eq:e14}
Y(a) \ \equiv \frac{a'}{a}P', \quad Z(a) \ \equiv (H_0)^2 \ P.
\end{equation}
These functions are reduced to
\begin{equation}
  \label{eq:e15}
Y(a) = a^{-5/2} (\Omega_M + \Omega_\Lambda a^3)^{1/2}  I(a), \quad 
Z(a) = \frac{2}{3\Omega_M} a [1 - Y(a)],
\end{equation}
where
\begin{equation}
  \label{eq:e16}
I(a) \equiv \int^a_0 \ db \ [b^3/(\Omega_M + \Omega_\Lambda b^3)]^{1/2}.
\end{equation}
\medskip
For $T_s$,  we assume the simplest transfer function (BBKS) for cold matter,
adiabatic fluctuations,  given by\citep{bbks}
\begin{equation}
  \label{eq:e17}
T_s (x) = \frac{\ln (1+0.171 x)}{0.171x} [1+0.284 x +(1.18 x)^2 + (0.399 x)^3 +
(0.490 x)^4]^{-1/4}.
\end{equation}
For  $x_{max}$ and $x_{min}$, we take  $x_{max} = 5.7$ and $x_{min} = 0.01$,
which were used in [I]. This value of $x_{max}$ 
corresponds to the lower limit of linear scales of super-horizon perturbations 
at the matter-dominant stage.\citep{t3} 

Then we obtain
\begin{equation}
  \label{eq:e18}
A = 2.22, \ \quad \ B = 20.95,
\end{equation}
$Y(1)$ and $Z(1)$ are shown in Eqs. (\ref{eq:b13}) and (\ref{eq:b14}) for the
 background parameter (\ref{eq:a2b}),  and 
\begin{equation}
  \label{eq:e19}
\langle  \mathop{\delta}_2 \rho/\tilde{\rho} \rangle = 0.121, 
\end{equation}
at the present epoch ($ a = 1$), where $\tilde{\rho} \equiv \rho + \Lambda$.

\section{Derivation of the deceleration parameter $q_{kin}$}

The background deceleration parameter $q$ is defined by
\begin{equation}
  \label{eq:f1}
q \equiv - \ddot{a} a/(\dot{a})^2.
\end{equation}
The corresponding parameter $q_{kin}$ is expressed as
\begin{equation}
  \label{eq:f2}
q_{kin} \equiv - \ddot{a}_{rem} a_{rem} /(\dot{a}_{rem})^2,
\end{equation}
where $a_{rem}$ is given using $a$ and $\langle l_{ii} \rangle$ in Eq.(\ref{eq:b6}).
Diffenretiaying $a_{rem}$, we obtain 
\begin{equation}
  \label{eq:f3}
      \begin{split}
\dot{a}_{rem} &= \dot{a} \Bigl(1 + \frac{1}{6} \langle l_{ii} \rangle \Bigr) + 
\frac{1}{6} a \langle l_{ii} \rangle^{.},  \\
\ddot{a}_{rem} &= \ddot{a} \Bigl(1 + \frac{1}{6} \langle l_{ii} \rangle \Bigr) + 
\dot{a} \frac{1}{3} \langle l_{ii} \rangle^{.} + \frac{1}{6} a \langle l_{ii} \rangle^{..}.
 \end{split}
\end{equation}
Using them, we obtain the Hubble parameter $H_{kin} \ (\equiv
 \dot{a}_{rem}/a_{rem})$ in Eq. (\ref{eq:b7}), and
\begin{equation}
  \label{eq:f4}
q_{kin} = q - \frac{1}{6 H^2}  \langle l_{ii} \rangle^{..} - \frac{1}{3} \frac{1+q}{H}
 \langle l_{ii} \rangle^{.}.
\end{equation}
Here we have the expression of $\langle l_{ii} \rangle^{.}$ in Eq.(\ref{eq:b9})
with $dZ(a)/da$.

Now we neglect the small terms with $A$, i.e.  
\begin{equation}
  \label{eq:f5}
\langle l_{ii} \rangle = \zeta  [Z(a)]^2, 
\end{equation}
and
\begin{equation}
  \label{eq:f6}
\langle l_{ii} \rangle^{.} = 2 \zeta \frac{Z(a) Y(a) \dot{a}}{\Omega_M +
 \Omega_\Lambda a^3},
\end{equation}
where 
\begin{equation}
  \label{eq:f6a}
\zeta \equiv  2 \pi \times 32.4^4 {\cal P}_{R0} B = 0.01523 \times 20.95
 = 0.319.
\end{equation}

First,  differetiating Eq.(\ref{eq:f5}), we get
\begin{equation}
 \label{eq:f7}
\begin{split}
\langle l_{ii} \rangle^{..} &=  2 \zeta  (\Omega_M + \Omega_\Lambda a^3)^{-1}
 Z(a) Y(a)  \\
&\times  \Bigl\{\ddot{a}  
 + (\dot{a})^2 \Bigl[\frac{dZ(a)/da}{Z(a)} + \frac{dY(a)/da}{Y(a)} -
  \frac{3\Omega_\Lambda a^2}{\Omega_M + \Omega_\Lambda a^3} \Bigr] \Bigr\}.
 \end{split}
\end{equation}
Using Eqs.(\ref{eq:b10}), (\ref{eq:e14}) and (\ref{eq:e15}),  we can derive
\begin{equation}
  \label{eq:f8}
- dY(a)/da = \frac{1}{a} \Bigl\{ \Bigl[1+ \frac{3}{2} \Omega_M/(\Omega_M +
 \Omega_\Lambda a^3) \Bigr] Y - 1 \Bigr\}.
\end{equation}
From Eqs.({\ref{eq:f7}) and ({\ref{eq:f8}), we obtain
\begin{equation}
  \label{eq:f9}
\langle l_{ii} \rangle^{..} =  2 \zeta  \frac{a H^2}{\Omega_M + 
\Omega_\Lambda a^3} \Bigl[- \Bigl(q + \frac{5}{2} + \frac{\frac{3}{2} 
\Omega_\Lambda a^3}{\Omega_M + \Omega_\Lambda a^3} \Bigr) \ Z \ Y
 + Z + \frac{a Y^2}{\Omega_M + \Omega_\Lambda a^3} \Bigr].
\end{equation}
Using Eqs.(\ref{eq:f4}) and (\ref{eq:f9}), we obtain
\begin{equation}
  \label{eq:f10}
q_{kin} = q - \frac{\frac{1}{3} \zeta a }{\Omega_M + 
\Omega_\Lambda a^3} \
\Bigl[\Bigl(q - \frac{1}{2} -  \frac{\frac{3}{2} \Omega_\Lambda a^3}
 {\Omega_M + \Omega_\Lambda a^3} \Bigr) \ Z \ Y 
 + Z + \frac{a Y^2}{\Omega_M + \Omega_\Lambda a^3} \Bigr].
\end{equation}
On the other hand, the background model gives
\begin{equation}
  \label{eq:f11}
q \equiv \Bigl(\frac{1}{2} \Omega_M - \Omega_\Lambda a^3 \Bigr)/(\Omega_M +
 \Omega_\Lambda a^3),
\end{equation}
where the present value is $q_0 = - 0.67$.

The present value of $q_{kin}$ is 
\begin{equation}
  \label{eq:f12}
(q_{kin})_0 - q_0 = -\frac{1}{3} \zeta  \ [- 3 \Omega_\Lambda Z(1) Y(1) + Z(1)
 + Y(1)^2].
\end{equation}

Using  the values of $Y(1)$ and $Z(1)$ in Eqs.(\ref{eq:b13}) and (\ref{eq:b14})
for the background model parameters, we obtain
\begin{equation}
  \label{eq:f16}
(q_{kin})_0 =  -0.670 + 0.011 = - 0.659.
\end{equation}
%
                      
\section{Derivation of $(d_L)_{rem}$}

Using the background equations for a light path, we obtain
\begin{equation}
  \label{eq:g1}
\begin{split}
a_{rem} (t_0) \ \  r &= \int^{t_0}_t dt \ a_{rem} (t_0)/a_{rem} (t)  \\
&= \int^1_a \frac{da}{a H_0} \ (\Omega_M  a^{-3} + \Omega_\Lambda)^{-1/2}
[1 + z_{rem} (a)],
 \end{split}
\end{equation}
where $z_{rem} (a)$ is expressed using Eq.(\ref{eq:b6}) as 
\begin{equation}
  \label{eq:g2}
1 + z_{rem} (a) = (1 + z) \Bigl\{1 + \frac{1}{6} [ \langle l_{ii} \rangle_0  -
\langle l_{ii} \rangle] \Bigr\}
\end{equation}
and $1 + z = 1/a(t)$. Using $(H_{kin})_0$ and Eq.(\ref{eq:d11}), therefore, 
$(d_L)_{rem}$ is expressed as
\begin{equation}
  \label{eq:g3}
(H_{kin})_0 \ (d_L)_{rem} = \frac{(H_{kin})_0}{H_0} (1+ z_{rem}) \times z W(a),
\end{equation}
where
\begin{equation}
  \label{eq:g4}
W(a) \equiv \frac{1}{z} \int^1_a \frac{da}{a^2} (\Omega_M  a^{-3} +
 \Omega_\Lambda)^{-1/2} \ \Bigl[1 + \frac{1}{6}  (\langle l_{ii} \rangle_0  -
\langle l_{ii} \rangle) \Bigr] .
\end{equation}
Using Eq.(\ref{eq:b5}), moreover, we can express $W(a)$ as
\begin{equation}
  \label{eq:g5}
W(a) =  \frac{1}{z} \int^1_a \frac{da}{a^2} (\Omega_M  a^{-3} +
 \Omega_\Lambda)^{-1/2} \Bigl\{1 + \frac{1}{6} \zeta \ [ Z(1)^2 - Z(a)^2] \Bigr\}.
\end{equation}
Here we have the relation for $z$ as
\begin{equation}
  \label{eq:g6}
z \ (= 1/a -1) = z_{rem} - \frac{1}{6} \zeta \ [ Z(1)^2 - Z(a)^2].
\end{equation}

For $z \ll 1$, we get
\begin{equation}
  \label{eq:g7}
z = z_{rem} - \frac{1}{3} \ \zeta \ \Bigl[ Z(a) \frac{Z(a)}{da} \Bigr]_{a=1} \ (1-a)
 = z_{rem} \ \Bigl[1 - \frac{1}{3} \zeta \ Z(1) Y(1) \Bigr] \ + O(z^2),
\end{equation}
where we used the relation $dZ(a)/da = Y(a)/(\Omega_M + 
\Omega_\Lambda a^3)$.
                      
On the other hand,  we have a relation
\begin{equation}
  \label{eq:g8}
\frac{(H_{kin})_0}{H_0} = 1 + \frac{1}{3} \zeta \ Z(1) Y(1), 
\end{equation}
neglecting small terms with $A$. So  we obtain finally
\begin{equation}
  \label{eq:g9}
(H_{kin})_0 \ (d_L)_{rem} \  = z_{rem} (1 + z_{rem}) \ \Phi (z_{rem}) \ W(a),  
\end{equation}
where $W(a)$ is given by Eq.(\ref{eq:g5}),  \  $a = 1/(1+z)$, and $z$ is related
 to $z_{rem}$ by Eq.(\ref{eq:g6}).     The auxiliary function $\Phi (z_{rem}) $
 is expressed as 
\begin{equation}
  \label{eq:g10}
\Phi (z_{rem})  \ \Bigl(\equiv \frac{(H_{kin})_0}{H_0} \frac{z}{z_{rem}} \Bigr) \ =
1+ \frac{1}{3}\zeta Z(1) Y(1) - \frac{1}{6} \zeta [Z(1)^2 - Z(a)^2]/z_{rem}.
\end{equation}

For $z_{rem} \ll 1$, \ $W$ and $\Phi$ are expanded as
\begin{equation}
  \label{eq:g11}
 (1 + z_{rem}) \ W = 1 + \Bigl[1 - \frac{3}{4} \Omega_M - \frac{1}{6} 
\zeta Z(1) Y(1) \Bigr]  \ z_{rem} +  \ \cdots,
\end{equation}
and
\begin{equation}
  \label{eq:g12}
 \Phi \ = \ 1 +  \frac{1}{6} \zeta \Bigl[\frac{1}{2} (1 - 3 \Omega_\Lambda )
 Z(1) Y(1)  + Y(1)^2 + \frac{3}{2}  \Omega_M  Z(1) ^2 \Bigr]  \ z_{rem} +  
 \ \cdots. 
\end{equation}

\bigskip
\section{Renormalized model parameters for  other background model parameters}

Let us show the renormalized model parameters for 
 other background model parameters such as
\begin{equation}
  \label{eq:h1}
H_0 = 67.3 \ {\rm km \ s^{-1} Mpc^{-1}}, \ \ \Omega_\Lambda = 1 - \Omega_M, 
\quad  {\rm{and}} \quad \Omega_M > 0.22.  
\end{equation}
In these cases, we have
\begin{equation}
  \label{eq:h2}
k_{eq} \ (\equiv a_{eq} H_{eq}) = 219 \ (\Omega_M h) = \ 147.4 \ \Omega_M,
\end{equation}
From Eqs. (\ref{eq:b12}) ,  (\ref{eq:c2}),  (\ref{eq:c4}) and (\ref{eq:e12}), we obtain
\begin{equation}
  \label{eq:h3}
(H_{kin})_0 /H_0 =  1 + \frac{2\pi}{3} \ (k_{eq})^4 \ {\cal P}_{R0} \ Y(1) Z(1) B, 
 \end{equation} 
\begin{equation}
  \label{eq:h4}
(H_{dyn})_0 /H_0 =  1 + \frac{2\pi}{3} \ (k_{eq})^4 \ {\cal P}_{R0} \ [1 - Y(1)] Z(1) B,
 \end{equation} 
\begin{equation}
  \label{eq:h5}
  (\Omega_\Lambda)_{rem} = \Omega_\Lambda \ [H_0/(H_{dyn})_0]^2 ,
 \end{equation} 
where we neglected  small terms with $A$. Here $Y(a)$ and $Z(a)$ 
depend on $\Omega_M$ and $\Omega_\Lambda$, while
$B$ does not depend on them, but on $x_{max}$ and $x_{min}$.
\bigskip

For $(\Omega_M, \Omega_\Lambda) = (0.24, 0.76)$, we have
\begin{equation}
  \label{eq:h6}
k_{eq} = 35.4, \quad Y(1) = 0.557, \quad Z(1) = 1.232,
 \end{equation} 
so that 
\begin{equation}
  \label{eq:h7}
  \begin{split}
(H_{kin})_0 /H_0 &=  \ 1 + 0.1043 \ (B/20.95), \\
(H_{dyn})_0 /H_0 &=  \ 1 + 0.0827 \ (B/20.95). 
\end{split}
\end{equation} 
Then  we obtain
\begin{equation}
  \label{eq:h8}
(H_{kin})_0 \ = \ 74.3, \quad  (H_{dyn})_0 \ = \ 72.8, \quad (\Omega_\Lambda)_{rem} = 
0.65
 \end{equation} 
for $B = 20.95 \ \ ({\rm with} \ x_{max} = 5.7)$. \ \ From Eqs.(\ref{eq:f11}) and 
(\ref{eq:f12}), moreover, we have
\begin{equation}
  \label{eq:h8a}
[q_0, \ (q_{kin})_0 ] = \ (-0.640, \ -0.638). 
\end{equation} 
\bigskip

For $(\Omega_M, \Omega_\Lambda) = (0.28, 0.72)$, we have

\begin{equation}
  \label{eq:h9}
k_{eq} = 41.3, \quad Y(1) = 0.540, \quad Z(1) = 1.095,
 \end{equation} 
so that 
\begin{equation}
  \label{eq:h10}
  \begin{split}
(H_{kin})_0 /H_0 &=  \ 1 + 0.166 \ (B/20.95), \\
(H_{dyn})_0 /H_0 &=  \ 1 + 0.141 \ (B/20.95).
\end{split}
\end{equation} 
Then  we obtain
\begin{equation}
  \label{eq:h11}
(H_{kin})_0 \ = \ 71.5, \quad  (H_{dyn})_0 \ = \ 70.8, \quad (\Omega_\Lambda)_{rem} = 
0.65
 \end{equation} 
for $B = 7.8 \ \ ({\rm with} \ x_{max} = 3.77)$. \ \ Moreover, we have
\begin{equation}
  \label{eq:h12}
[q_0, \ (q_{kin})_0 ] = \ (-0.580, \ -0.568).   
 \end{equation} 
\bigskip

 \bigskip

\bigskip
    

\end{document}